\documentclass[aps,prd,twocolumn,superscriptaddress,showpacs, showkeys]{revtex4-1}
\usepackage[utf8]{inputenx} 
\usepackage[english]{babel} 
\usepackage{subfigure} 
\usepackage{hyperref} 
\usepackage{graphicx}
\usepackage{amsmath,amssymb}
\usepackage{lineno}
\usepackage{comment}
\usepackage{overpic}

\newcommand{\DBD}{0$\nu$DBD}

\newcommand{\TEO}{$\mathrm{TeO}_2$}
\newcommand{\TEHT}{$^{130}\mathrm{Te}$}

\newcommand{\CUORICINO}{CUORICINO}
\newcommand{\CUORE}{CUORE}

\newcommand{\CUOREZ}{CUORE-0}

\newcommand{\TEU}{$^{123}\mathrm{Te}$}
\newcommand{\MC}{Monte Carlo}

\providecommand*{\un}[1]{\ensuremath{\mathrm{~#1}}}

\begin{document}       

\title{The low energy spectrum of TeO$_2$ bolometers: results and dark matter perspectives for the CUORE-0 and CUORE experiments}
\author{F.~Alessandria}
\affiliation{INFN - Sezione di Milano, Milano I-20133 - Italy}

\author{R.~Ardito}
\affiliation{Dipartimento di Ingegneria Strutturale, Politecnico di Milano, Milano I-20133 - Italy}

\author{D.~R.~Artusa}
\affiliation{Department of Physics and Astronomy, University of South Carolina, Columbia, SC 29208 - USA}
\affiliation{INFN - Laboratori Nazionali del Gran Sasso, Assergi (L'Aquila) I-67010 - Italy}

\author{F.~T.~Avignone~III}
\affiliation{Department of Physics and Astronomy, University of South Carolina, Columbia, SC 29208 - USA}

\author{O.~Azzolini}
\affiliation{INFN - Laboratori Nazionali di Legnaro, Legnaro (Padova) I-35020 - Italy}

\author{M.~Balata}
\affiliation{INFN - Laboratori Nazionali del Gran Sasso, Assergi (L'Aquila) I-67010 - Italy}

\author{T.~I.~Banks}
\affiliation{Department of Physics, University of California, Berkeley, CA 94720 - USA}
\affiliation{Nuclear Science Division, Lawrence Berkeley National Laboratory, Berkeley, CA 94720 - USA}
\affiliation{INFN - Laboratori Nazionali del Gran Sasso, Assergi (L'Aquila) I-67010 - Italy}

\author{G.~Bari}
\affiliation{INFN - Sezione di Bologna, Bologna I-40127 - Italy}

\author{J.~Beeman}
\affiliation{Materials Science Division, Lawrence Berkeley National Laboratory, Berkeley, CA 94720 - USA}

\author{F.~Bellini}
\affiliation{Dipartimento di Fisica, Sapienza Universit\`a di Roma, Roma I-00185 - Italy }
\affiliation{INFN - Sezione di Roma, Roma I-00185 - Italy }

% ...

\begin{abstract}
We collected 19.4 days of data from four 750~g TeO$_2$ bolometers, and in
three of them we were able to set the energy threshold around 3~keV using a
new analysis technique. We found a background rate ranging from 
25 cpd/keV/kg at 3~keV to 2~cpd/keV/kg at 25~keV, and a peak at 4.7
keV. The origin of this peak is
presently unknown, but its presence is confirmed by a reanalysis of 62.7
kg$\cdot$days of data from the finished \CUORICINO\ experiment. Finally,
we report the expected sensitivities of the \CUOREZ\ (52 bolometers) and
\CUORE\ (988 bolometers) experiments to a WIMP annual modulation signal.
\end{abstract}
\maketitle

\section{Introduction}

Tellurium dioxide bolometers are excellent detectors to search
for rare processes.  Operated at a temperature of about 10\un{mK}, these
detectors feature an energy resolution of a few keV over an energy range
extending from a few keV up to several MeV.  This, together with the
low level of radioactive background achievable and the low cost, makes
them ideal detectors for \CUORE, an experiment that will search
for neutrinoless double beta decay (\DBD) of \TEHT~\cite{Arnaboldi:2003tu,ACryo}.

\CUORE\ will consist of 988 \TEO\ bolometers of 750\un{g} each and is expected to
reach a background level at the \TEHT\ Q-value (around 2528 keV~\cite{Redshaw:2009zz,scielzo09,Rahaman:2011zz}) of
the order of 0.01\un{counts/keV/kg/y}, thus allowing a sensitivity to \DBD\ down
to the inverted hierarchy of neutrino masses. \CUORE\
is currently being installed at the Laboratori Nazionali del Gran Sasso
(LNGS) in Italy and is scheduled to start taking data in 2014. The
experimental design was validated by \CUORICINO, an array of 62 \TEO\
bolometers (for total mass of 40.7\un{kg}) which took data at LNGS from 2003 
to 2008 and put stringent limits on the \DBD\ 
half-life~\cite{Andreotti:2010vj,Andreotti:2011in}.  To test the new assembly line
and the materials chosen for \CUORE, an array of 52 \TEO\ bolometers, 
named \CUOREZ, has been recently built. It is expected to have a smaller background than \CUORICINO\ and will start taking data in Fall 2012. 

Given the high mass, the good energy resolution, and the low background,
\CUOREZ\ and \CUORE\ experiments can search for rare processes such
as Dark Matter interactions.  While there is significant evidence
that Dark Matter exists~\cite{Bertone:2004pz}, its composition is
as of yet unknown. Weakly Interacting Massive Particles (WIMPs) are
the theoretically favored candidates~\cite{Steigman:1984ac},  and
several earthbound experiments have been designed to detect their
scattering off nuclei, which should produce energy releases of a few
keV~\cite{Goodman:1984dc}. Many experiments use multiple signatures
(e.g. phonon, scintillation and ionization) to discriminate nuclear
recoils from the $\beta/\gamma$ background originating from natural
radioactivity. Other experiments search for the expected annual modulation
of the interaction rate as the Earth traverses the uniform Dark Matter
distribution in the galactic halo~\cite{Drukier:1986tm}. Some experiments
have reported compatibility with the signal expected for a galactic halo
WIMP population, but have been met with skepticism~\cite{pdg2012}. As
our knowledge of the properties of Dark Matter is limited, a variety of
different experimental approaches are required and the search for Dark
Matter with \TEO\ bolometers could provide important new information.

Such searches were not performed in \CUORICINO\ because
the energy threshold was too high, of the order of tens of keV. A new software trigger 
was developed to lower the energy threshold~\cite{DiDomizio:2010ph}. 
The trigger  is based on the matched filter algorithm~\cite{Gatti:1986cw,Radeka:1966}
and also provides a pulse shape parameter to suppress false signals
generated by detector vibrations and electronics noise.

In this paper we show the energy spectrum of four \TEO\ bolometers originally
operated at LNGS to test the performances
 of \CUORE\ crystals~\cite{Alessandria:2011vj}. In three of them we were
 able to set the energy threshold around 3\un{keV}, while the fourth was set to 10\un{keV} 
 because of higher detector noise.
The energy spectrum of the three bolometers exhibits a peak at 4.7\un{keV} whose origin
is presently unknown.  The peak has a constant rate in time, and
its presence is confirmed by a reanalysis of the data from the last two
months of operation of \CUORICINO.  
Given the observed counting rate,
we also evaluate the sensitivity of \CUOREZ\ and \CUORE\ to an annual
modulation signal induced by  WIMP Dark Matter candidates, comparing it to the results of other experiments.
 
\section{Experimental setup} 

A \CUORE\ bolometer is composed of two main parts: a \TEO\
crystal and a neutron transmutation doped Germanium (NTD-Ge)
thermistor~\cite{wang,Itoh}. The crystal, a $750\un{g}$ cube of
side length $5\un{cm}$, is held by PTFE supports in copper frames.
The frames are connected to the mixing chamber of a dilution
refrigerator, which keeps the bolometers at a temperature of about
$10\un{mK}$.  The thermistor is glued to the crystal and acts as
thermometer.  The temperature increase due to an energy deposition
in the crystal is measured by the decrease in the resistance of the
thermistor~\cite{Mott:1969,Itoh:1996}.  The thermistor is biased
by a constant current and a change in the voltage across it constitutes
the signal~\cite{AProgFE}.  In the setup presented in this paper, called
\CUORE\ Crystal Validation Run 2 (CCVR2), each crystal
was provided with two thermistors for redundancy. 
Two Joule heaters  were also glued on two crystals and used to inject heat 
pulses with controlled amounts of energy to monitor the detector gain 
and efficiencies~\cite{stabilization,Arnaboldi:2003yp,Andreotti:2012zz}.  
The bolometers were operated in the cryogenic R\&D facility of \CUORE\ 
whose details can be found in Refs.~\cite{Pirro:2006mu,Arnaboldi:2006mx,Arnaboldi:2004jj}.

Data were collected over 23\un{days} with small interruptions for
calibrations and cryostat maintenance. The effective live time amounted
to 19.4\un{days}.  On each of the four bolometers, labeled B1, B2,
B3 and B4, we selected the thermistor in which the trigger reached the
lowest energy threshold.

The energy calibration was performed by inserting two wires of thoriated
tungsten in proximity of the detector, between the cryostat and the
lead shields placed externally. The main $\gamma$ lines of $^{232}$Th,
ranging from 511 to 2615\un{keV}, and the lines generated by metastable
Te isotopes in the crystals, ranging from 30 to 150\un{keV}, were used
for the determination of the calibration function. These isotopes are
produced by cosmogenic activation during production and
air-shipment of the crystals outside the underground laboratory.

\section{Detector performance}

The energy resolution was evaluated both at the baseline level and on
the lowest energy peak in the spectrum. Randomly triggered events,
not containing pulses, were used to evaluate the fluctuation of the
baseline after the application of the matched filter.  A Gaussian fit
was performed on the 30.5\un{keV} Sb line, which is due to EC decays
of metastable Te isotopes.  As one can
see from Tab.~\ref{tab:OTeff},  baseline and 30.5\un{keV} resolutions
are quite close, since at these energies the resolution of bolometers is
expected to be dominated by the noise. The bolometers feature excellent 
energy resolution, with B2, B3, and B4 below 1\un{keV~FWHM}, and B1 
at 3.3\un{keV~FWHM}.

The detection efficiencies were measured via an energy scan performed
at the end of the run by using the Joule heaters to provide
sequences of pulses from 1 to $50\un{keV}$.  However,
heater events and actual particle events have slightly different pulse
shapes which could result in slightly different energies and detection
efficiencies.  
The trigger efficiency and any differences that may result from the difference between the heater pulses and particle events were investigated by \MC\ simulations.

\begin{figure}[htb]
\centering \includegraphics[width=0.49\textwidth]{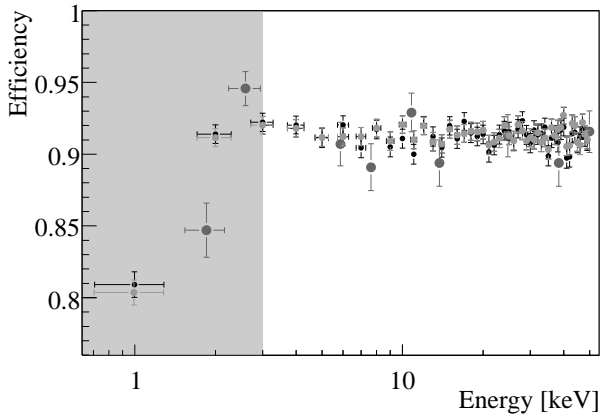}
\caption{Detection efficiency on B2: particle \MC\ in black circles,
heater \MC\ in green circles, heater scan data in big red circles. The
shadowed band represents the region below the energy threshold chosen for
the data analysis.}
\label{fig:MCsignal3}
\end{figure}

Heater pulses between 1--50\un{keV} and particle pulses in the same energy range were simulated separately using the tools described in Ref.~\cite{Carrettoni:2011rn}.  Only B2 and B3 were simulated for heater pulses while all the four bolometers were considered for the particle pulses simulation.
Both \MC\ runs included simulated
detector noise and the background from particle pulses, randomly generated
to match the event rate and the entire measured energy spectrum  from threshold
up to 5.4\un{MeV}, i.e. the energy of $^{210}$Po decays
in the crystal.  We processed the output of the simulation in the same
way as the heater scan measurement.  The estimated efficiencies on B2
are shown in Fig.~\ref{fig:MCsignal3}, where the good agreement
between \MC\ and data is evident.  

The similarities in the energy dependence demonstrate that the trigger
acts in the same way on particle and heater pulses. This feature
is visible also on B3, the other bolometer with a heater.  We 
estimate the detection efficiencies on the two bolometers without heater
(B1 and B4) using the \MC.  For each bolometer, we set the energy
threshold for the data analysis as the energy at which the plateau
is reached. The detection efficiency $\epsilon_D$ is computed as the
weighted average of each point in the plateau and it is considered constant
in the data analysis above threshold. The energy threshold,
the heater-measured, and particle-simulated detection efficiencies of
each bolometer are reported in Tab.~\ref{tab:OTeff}. 
The heater-simulated detection efficiencies
are found consistent with the heater-measured ones within $1 \sigma$.

The residual 10-20\% inefficiency in the plateau is due to the trigger dead time, which is mainly due to $^{210}$Po decays. $^{210}$Po is introduced in the
crystal growth and has a half-life of 147 days. For sufficiently aged crystals, such as the \CUOREZ\ and \CUORE\ ones, the dead time is expected to be negligible because this activity has decayed away, raising the efficiencies up to 100\% (see Ref.~\cite{DiDomizio:2010ph} for further details).
\begin{table}[htb]
\begin{center}
\caption{Baseline resolution ($\Delta E_{\rm base}$), resolution at 30.5\un{keV} ($\Delta E_{30\un{keV}}$), 
software energy threshold ($\theta_E$) and detection
efficiencies measured with the Joule heater ($\epsilon_D^{heater}$) and estimated from the \MC\
simulation of the four bolometers ($\epsilon_D^{MC}$).} \label{tab:OTeff}
\begin{tabular}{|c|c|c|c|c|c|}
\hline
Bolo& $\Delta E_{\rm base}$ & $\Delta E_{30\un{keV}}$ & $\theta_E$ & $\epsilon_D^{heater}$ & $\epsilon_D^{MC} $ \\
          & \multicolumn{2}{|c|}{[keV FWHM]}       & [keV]  &&\\
\hline
B1 & $ 3.3 $   &  $3.3 \pm 0.7$   & 10.0 & n/a & $0.878 \pm 0.002$  \\
B2 & $ 0.66 $  &  $0.53 \pm 0.08$ & 3.0 &  $0.910 \pm 0.005$ & $0.913 \pm 0.001$   \\
B3 & $0.76 $   &  $0.83 \pm 0.09$ & 2.5 & $0.828 \pm 0.007$ & $0.825 \pm 0.001$  \\
B4 & $0.82 $   &  $0.59 \pm 0.12$ & 2.5 & n/a & $0.828 \pm 0.002$  \\
\hline
\end{tabular}
\end{center}
\end{table}

\section{Energy calibration}

\begin{figure}[b] \centering
\begin{overpic}[width=0.49\textwidth]{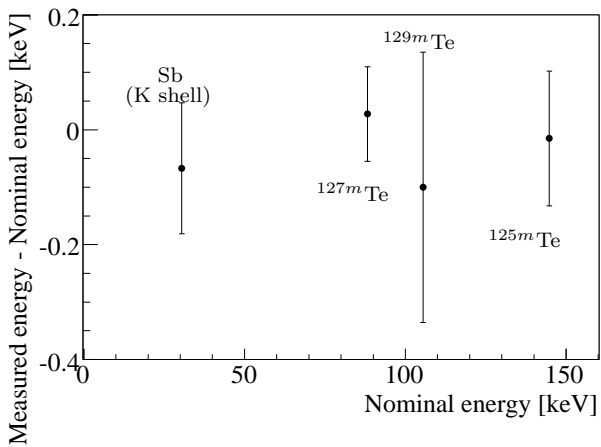}
\put(26,55){\footnotesize Sb}
\put(21,52){\footnotesize (K shell)}
\put(50,37){\footnotesize $^{127m}$Te}
\put(60,60){\footnotesize $^{129m}$Te}
\put(76,30){\footnotesize $^{125m}$Te}
\end{overpic}
\caption{Residuals of the calibration function of B2 on metastable Te lines.} 
\label{fig:metares} 
\end{figure}

The calibration function is a third order polynomial without intercept. 
The residual with respect to the nominal energy of the
peaks in the low energy spectrum is shown in Fig.~\ref{fig:metares}.

\begin{figure}[tbp]
\centering 
\includegraphics[width=0.49\textwidth]{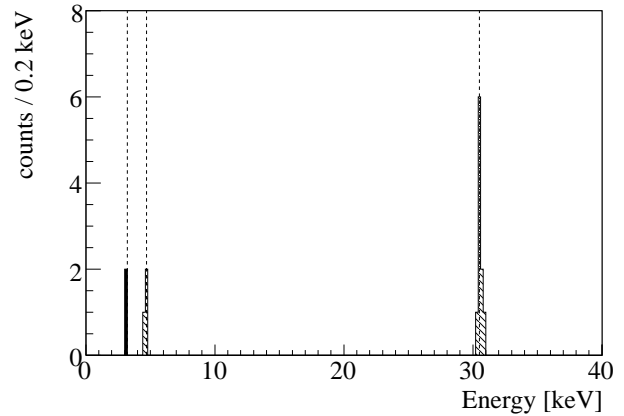}
\caption{Summed energy spectrum from B2, B3 and B4 in time coincidence with 
an energy deposition on another bolometer (including B1)  compatible
with the $\gamma$ lines from $^{40}$K (solid) or $^{121}$Te decays
(hashed).  The dashed lines represent the nominal values of the
de-excitation energies. No event is observed away from these lines.} 
\label{fig:coincidences}
\end{figure}

The accuracy at energies lower than $30\un{keV}$ was
verified by using events from $^{121}$Te and $^{40}$K contaminations
in the crystals.  These isotopes may decay via EC to $^{121}$Sb and
$^{40}$Ar, respectively, with the emission of a $\gamma$-ray from the
daughter nucleus de-excitation ($507.6\un{keV}$ or $573.1\un{keV}$
 from $^{121}$Te and 1461\un{keV} from $^{40}$K) and the de-excitation
 of the atomic shell L$_1$ and K ($E_{L1}=4.6983\un{keV}$
or $E_K=30.4912\un{keV}$ from $^{121}$Te and $E_K=3.2060\un{keV}$
from $ ^{40}$K). The atomic de-excitation process is fully contained
in the crystal while, in some cases, the $\gamma$ can escape and hit
another crystal, thus producing a double hit event. To select double
hit events, we set the time coincidence window to 10~\un{ms}, i.e. the 
measured jitter between the response of the bolometers.  We required that in one of the two crystals there is an energy
release corresponding to the $\gamma$-rays from $^{121}$Te or $^{40}$K,
and then, after applying the SI cut, we selected energy depositions between threshold and 40\un{keV} in the other crystal (Fig.~\ref{fig:coincidences}). Two events from  $^{40}$K
were found (3.04 and 3.18\un{keV}), three from the L$_1$ (4.48, 4.67 and
4.74\un{keV}), and ten from the K de-excitation after the $^{121}$Te
decay ($30.53\pm 0.04$\un{keV}). Since all the events are compatible
with their expected energy, the accuracy of the low-energy calibration
is confirmed.  We also note that the K/L capture ratio for $^{121}$Te
is compatible within two standard deviations with the expected value
of 7~\cite{Ohya20101619}.

\section{Event selection}

\begin{figure}[bt]
\centering
\begin{overpic}[width=0.49\textwidth]{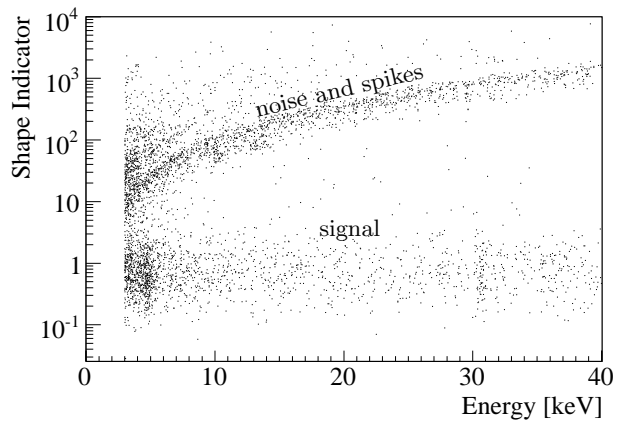}
\put(40,50){\rotatebox{13}{noise and spikes}}
\put(50,32){signal}
\end{overpic}
\caption{Pulse shape indicator (SI) distribution on B2.  The band at low
SI is populated by signal events.  At higher values there are triggered
mechanic vibrations, electronic spikes and pileups. The optimal cut for
this bolometer was found to be ${\rm SI}<2.2$.} \label{fig:scatterSI}
\end{figure}

We selected signal events using the shape indicator variable (SI)
described in Ref.~\cite{DiDomizio:2010ph}. This variable is based 
on the $\chi^2$ of the fit of the waveforms with the expected shape of the signal.
The distribution of SI versus energy for B2 is shown in Fig.~\ref{fig:scatterSI}. 
The signal region is easily identified, and a considerable part of the pulses generated
by thermal and electronic noise
can be removed with a cut on this variable.

We evaluated the optimal value of the cut on SI using the Sb K line
at $30.5\un{keV}$.  We performed a series of extended
unbinned maximum-likelihood fits, selecting the events below a fixed SI value.  For each
SI cut, the spectra of cut accepted events and cut rejected events were
simultaneously fitted with a Gaussian,  representative of the signal,
plus a first-order polynomial function, representative of the background.
Signal ($\epsilon_S$) and background ($\epsilon_B$) efficiencies after the cut
were evaluated directly in the fit. For each bolometer we selected
the cut which maximized the statistical significance, defined as
$\epsilon_S/\sqrt{\epsilon_B}$, corresponding in this case to
$\epsilon_S=1$. 

No anti-coincidence cut is applied. Since over the entire energy range multiple-hit events  
are mainly due to random coincidences, the application of anti-coincidence
cuts reduces the statistics without gaining in background reduction.

\section{Measured spectra}

\begin{figure}[b]
\centering 
\begin{overpic}[width=0.23\textwidth,clip=true]{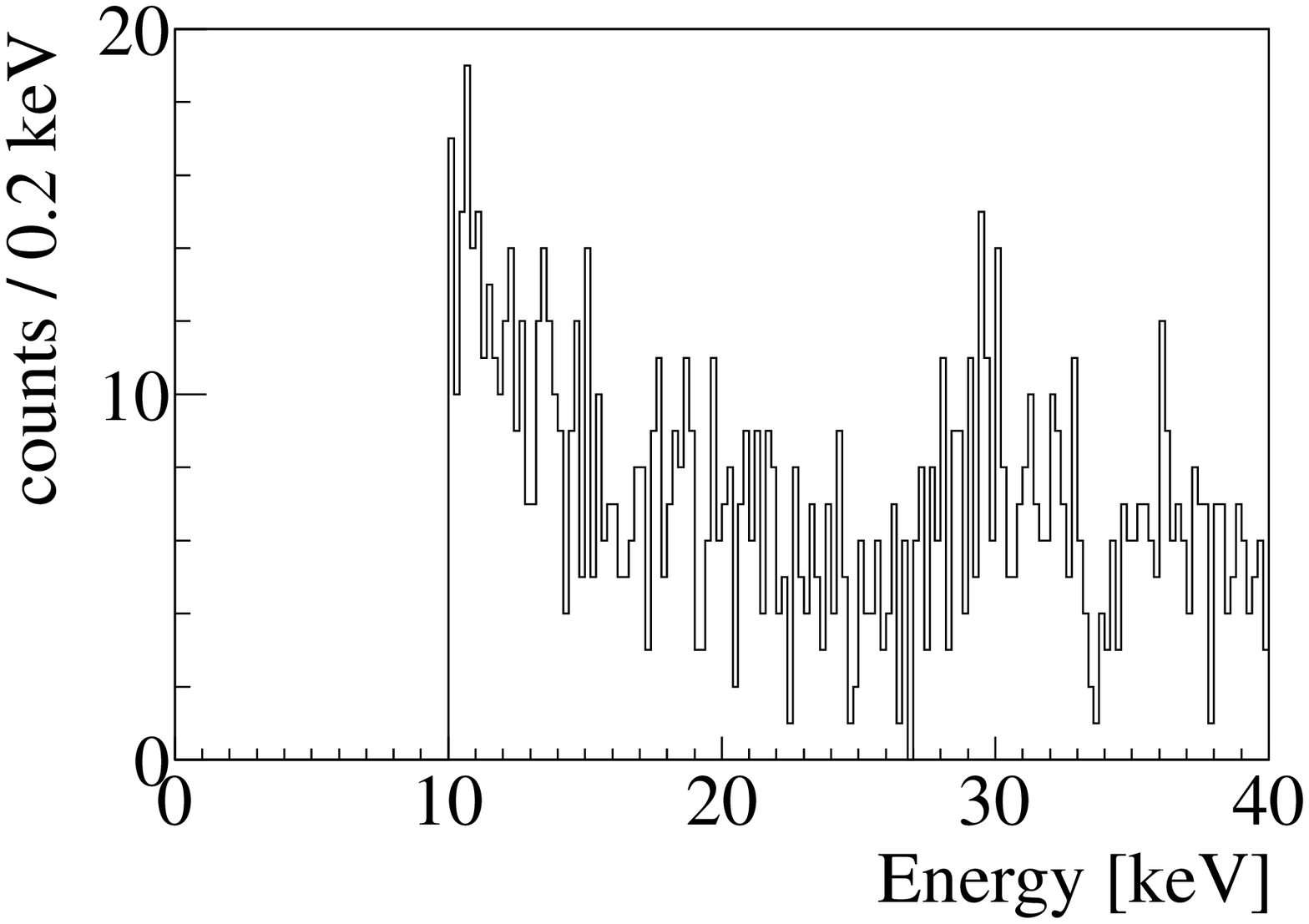}
\put(79,58){B1}
\end{overpic}
\begin{overpic}[width=0.23\textwidth,clip=true]{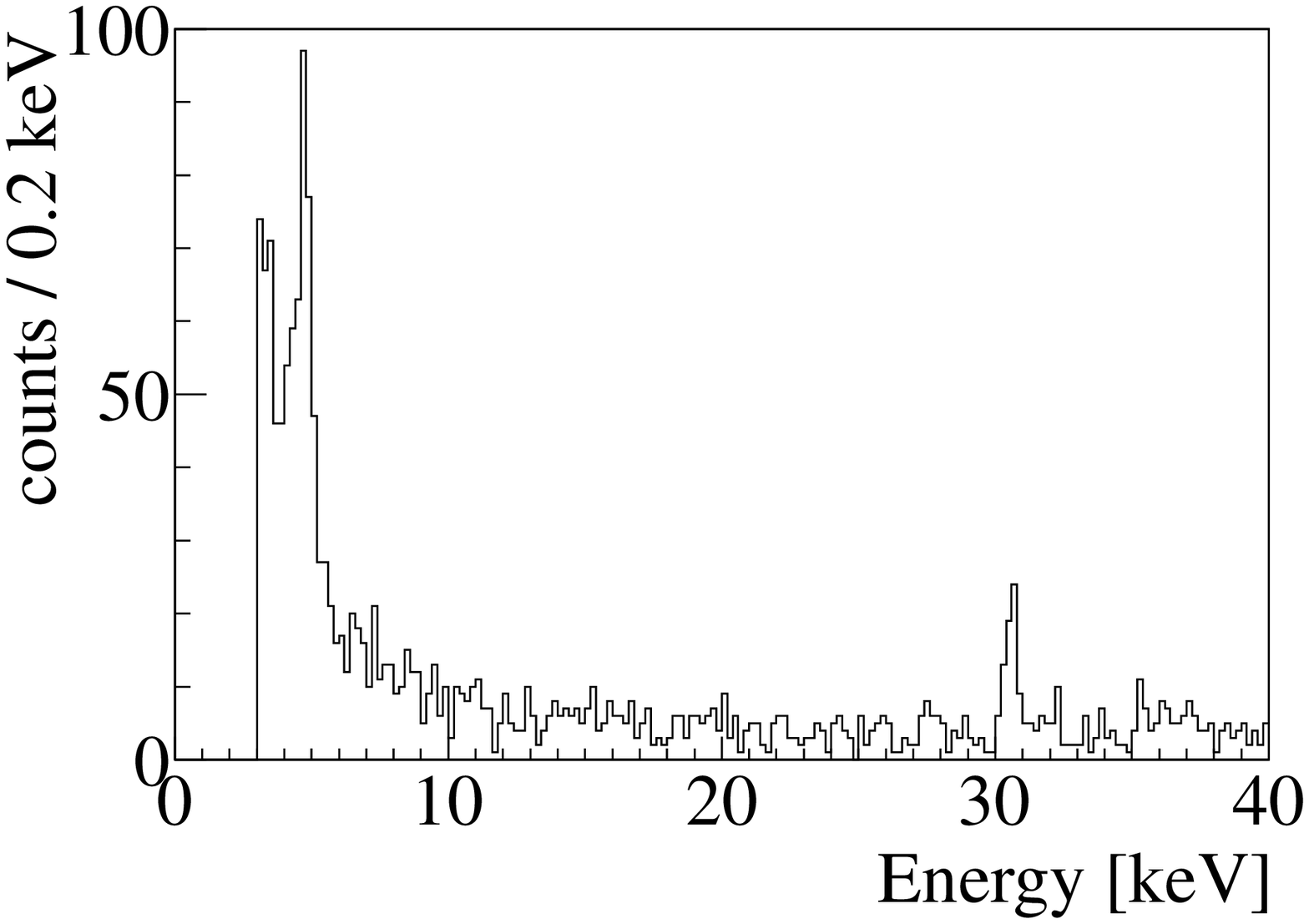}
\put(79,58){B2}
\end{overpic}
\begin{overpic}[width=0.23\textwidth,clip=true]{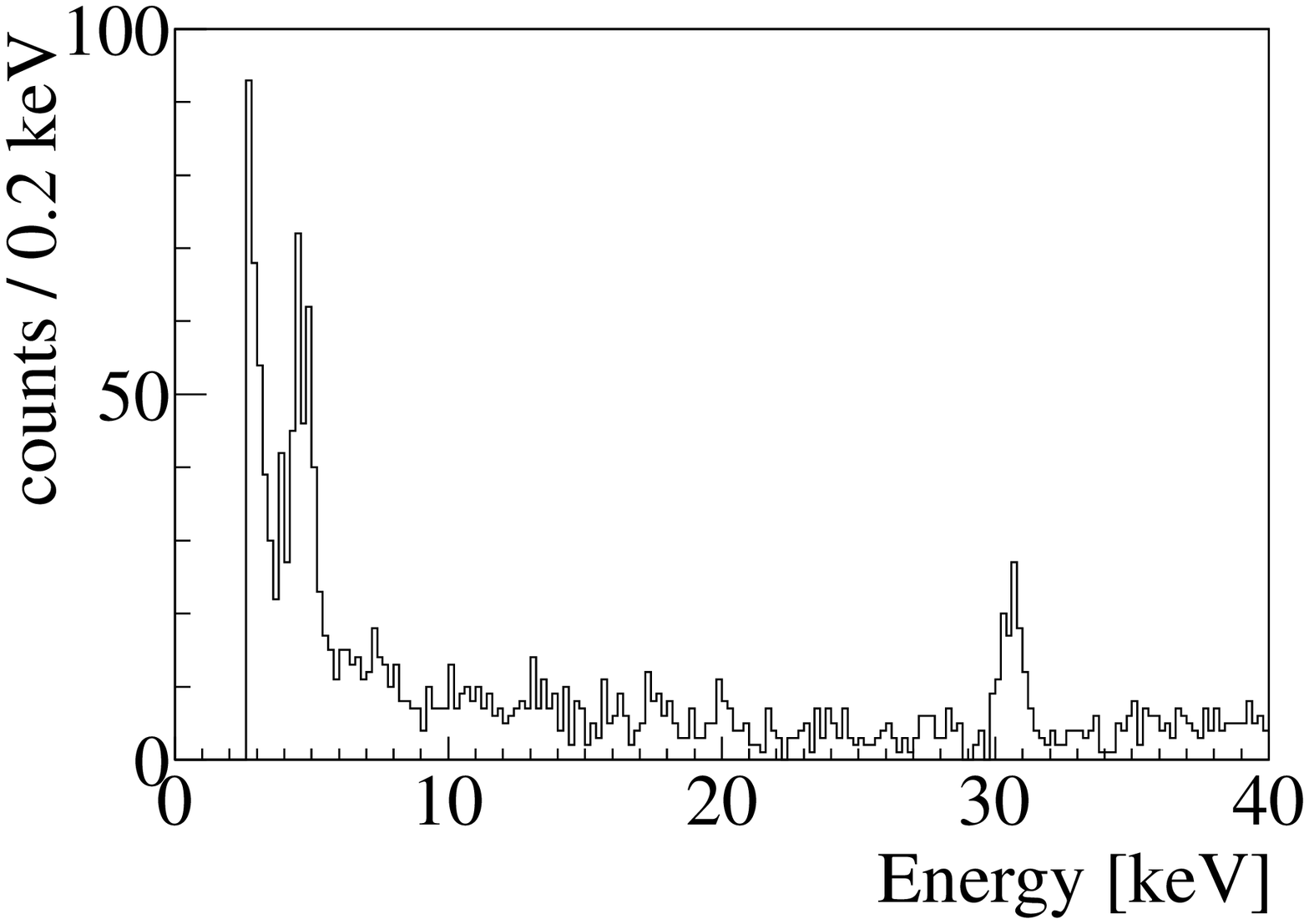}
\put(79,58){B3}
\end{overpic}
\begin{overpic}[width=0.23\textwidth,clip=true]{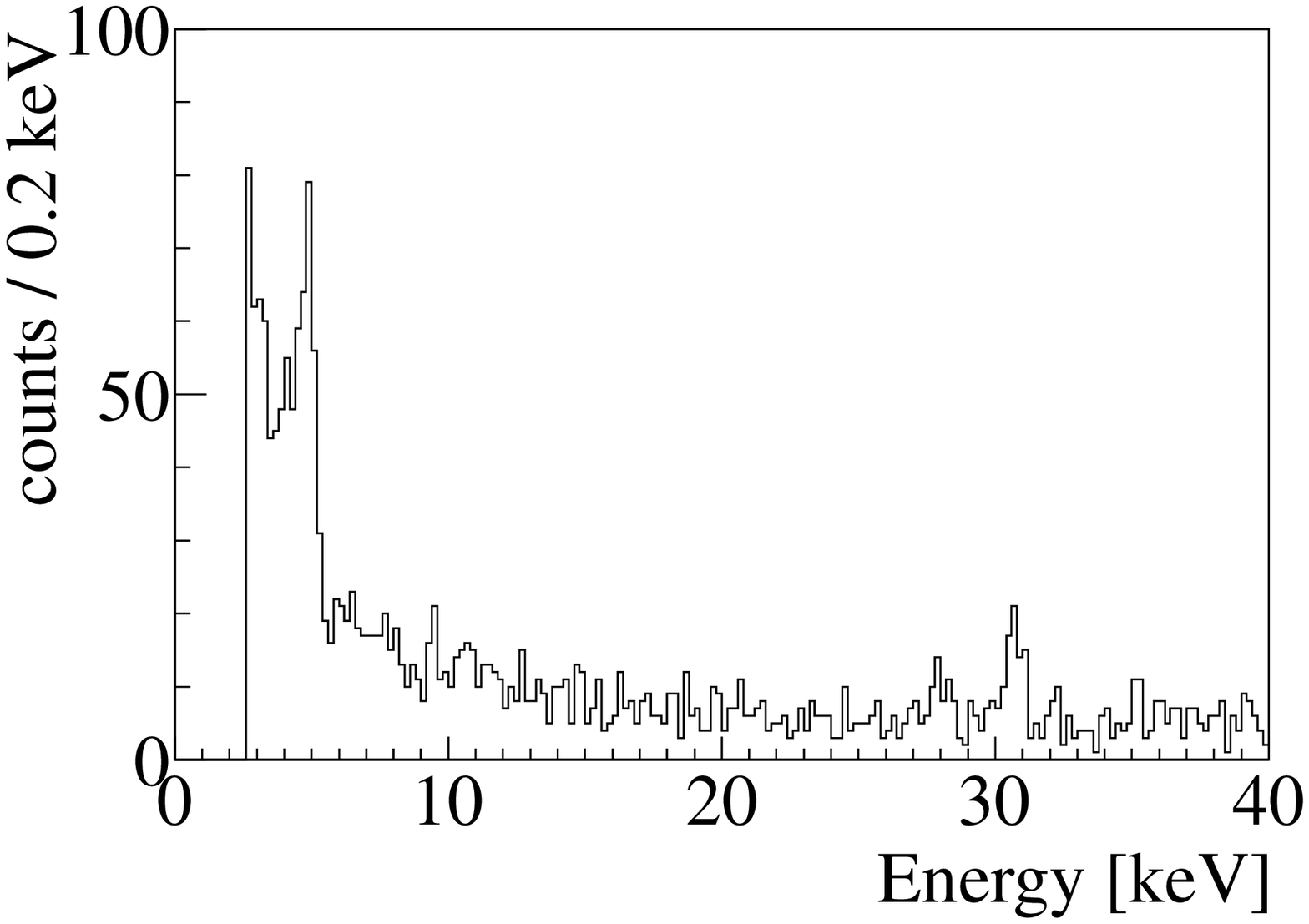}
\put(79,58){B4}
\end{overpic}
\caption{Low energy spectra of the four bolometers. Full statistics
(19.4 days) with pulse shape cut applied.} \label{fig:spectra_best}
\end{figure}

The energy spectra of the four detectors show several $\gamma$ and $\alpha$
peaks that are clearly identified as due to U, Th and K contaminations
(of the crystals themselves and of the experimental setup) and few low
energy peaks identified as being due to Te metastable isotopes. 
The region between
threshold and $40\un{keV}$ is shown in Fig.~\ref{fig:spectra_best}.
In the three bolometers with the lowest threshold, a peak appears at
about $4.7\un{keV}$, the energy of the L$_1$ atomic shell of Sb. This
in principle indicates that the line could be ascribed to the EC decay
of a Te isotope, however none of the known or predicted isotope decays 
can explain our observation:
\begin{itemize}
\item The $^{121m}$Te and its daughter $^{121}$Te  decay via EC with
half-lives of 154 and 17 days, respectively.  However the  intensity of
the observed peak is higher than the K line of Sb at 30.5 keV. This is
in contradiction with the measured value of the K/L$_1$ capture ratio for
these isotopes which is greater than one.
\item The other EC decaying metastable Te isotopes (and their daughters) have half-lives smaller than 4.7\un{days}. As it will demonstrated later,
the observed peak rate is constant over 20\un{days}.
\item \TEU\ is a naturally occurring isotope of Tellurium (abundance  $0.908\pm0.002\,\%$~\cite{ohya2004nuclear})  
which may decay via EC to $^{123}$Sb with a Q-value of $52.2 \pm 1.5
\un{keV}$.  Since the transition is 2$^{\rm nd}$-forbidden unique,
it has been estimated that very little K capture actually occurs
and that the majority of EC decays take place from the L$_3$  shell.
Several searches for Sb K lines from EC of \TEU\ have been performed.
Positive evidence, $T_{1/2} = (1.24 \pm 0.10) \cdot 10^{13}\un{y}$, was
claimed~\cite{watt1962search} but subsequently ruled
out~\cite{zuber:2003, PhysRevC.67.014323}.  In this paper, we show
for the first time an energy spectrum down to the L energy region. The
line we observe however cannot be attributed to $^{123}$Te, since, 
as described in the next section, the
energy is compatible with the L$_1$ shell (4.6983\un{keV}) and not with
the L$_3$ one (4.1322\un{keV}).

\end{itemize}

To further investigate the peak origin, in the next future we will operate
crystals enriched in $^{128}$Te and $^{130}$Te (therefore depleted in
other Te isotopes) and see if the peak intensity changes. In the following we
report the analysis of the peak in the energy distribution of single
bolometers, to provide all the possible details and to stimulate a
discussion within the scientific community that will hopefully lead to its
identification. 

To determine the intensity, the peaking background due to
the EC decay of $^{121m}$Te--$^{121}$Te was removed. The $\gamma$-rays
produced by these isotopes can escape without hitting another crystal,
such that only the K or the L de-excitations are measured. The number of
pure L events from $^{121}$Te and $^{121m}$Te ($N_{121}$) were estimated
using a \MC\ simulation based on GEANT4~\cite{allison2006geant4},
normalizing the single hit spectra to the measured rate of the most
intense $^{121m}$Te peak (294.0\un{keV}).

\begin{figure}[tbp]
\centering 
\begin{overpic}[width=0.23\textwidth,clip = true]{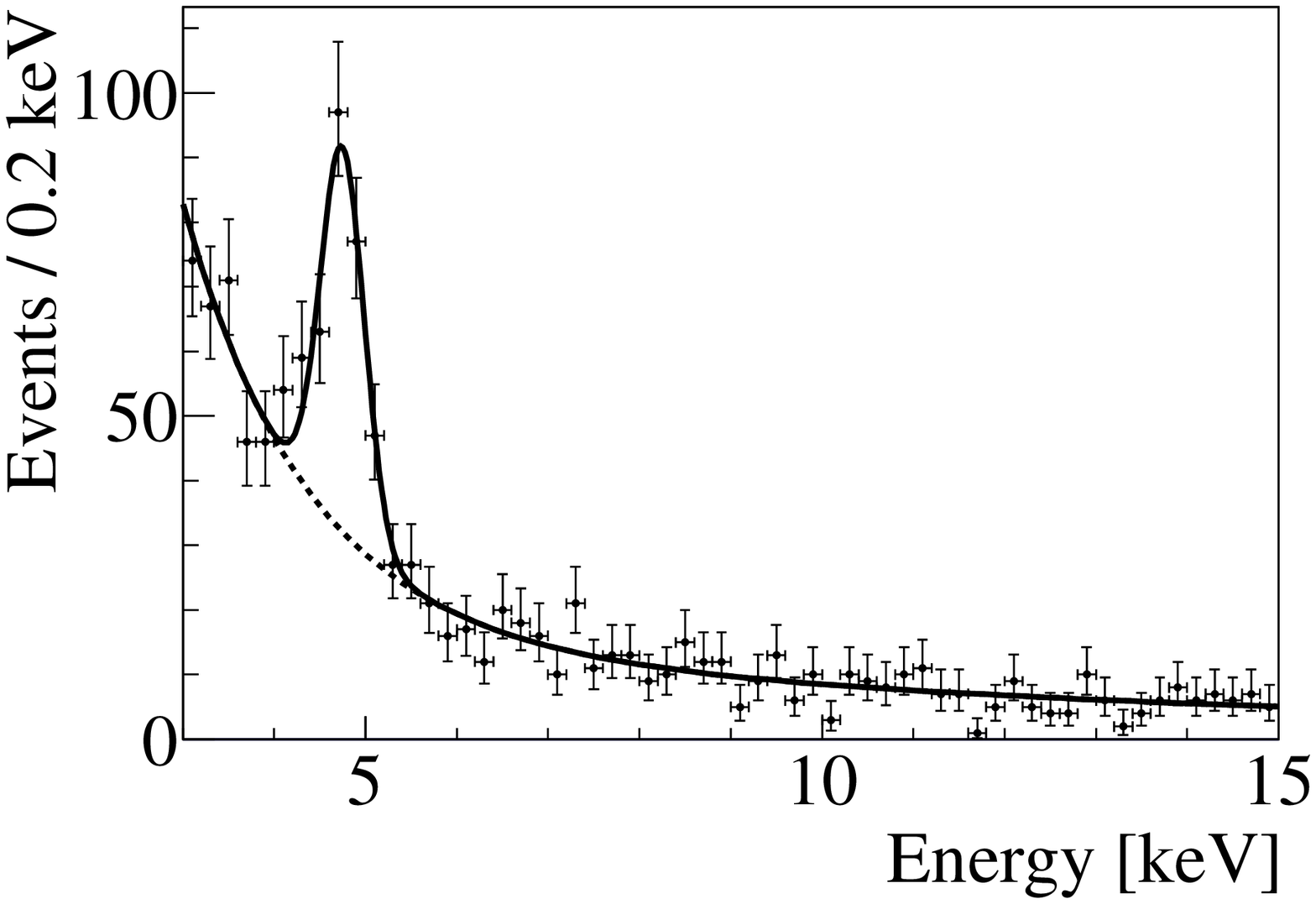}
\put(79,58){B2}
\end{overpic}
\begin{overpic}[width=0.23\textwidth,clip = true]{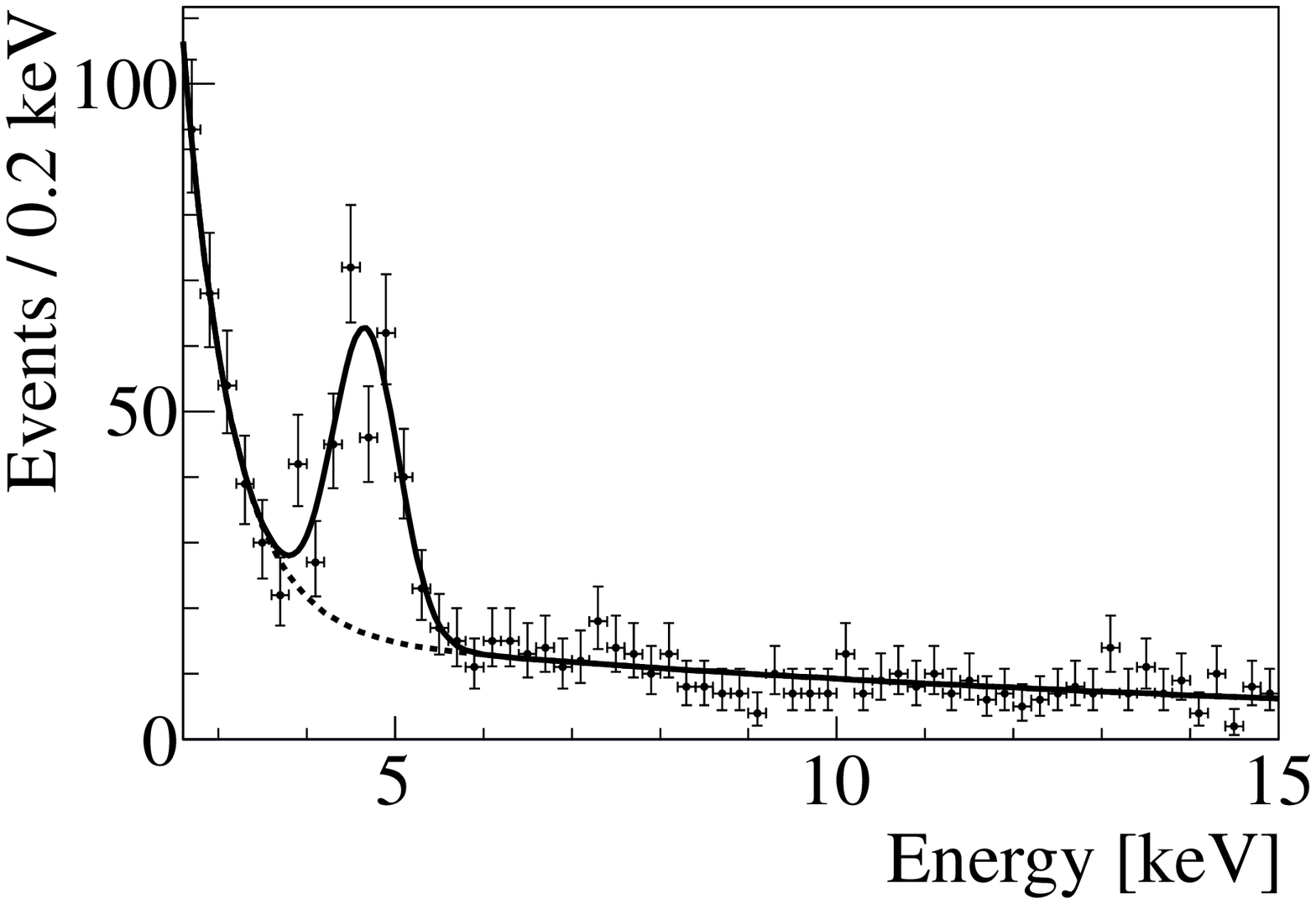}
\put(79,58){B3}
\end{overpic}
\begin{overpic}[width=0.23\textwidth,clip = true]{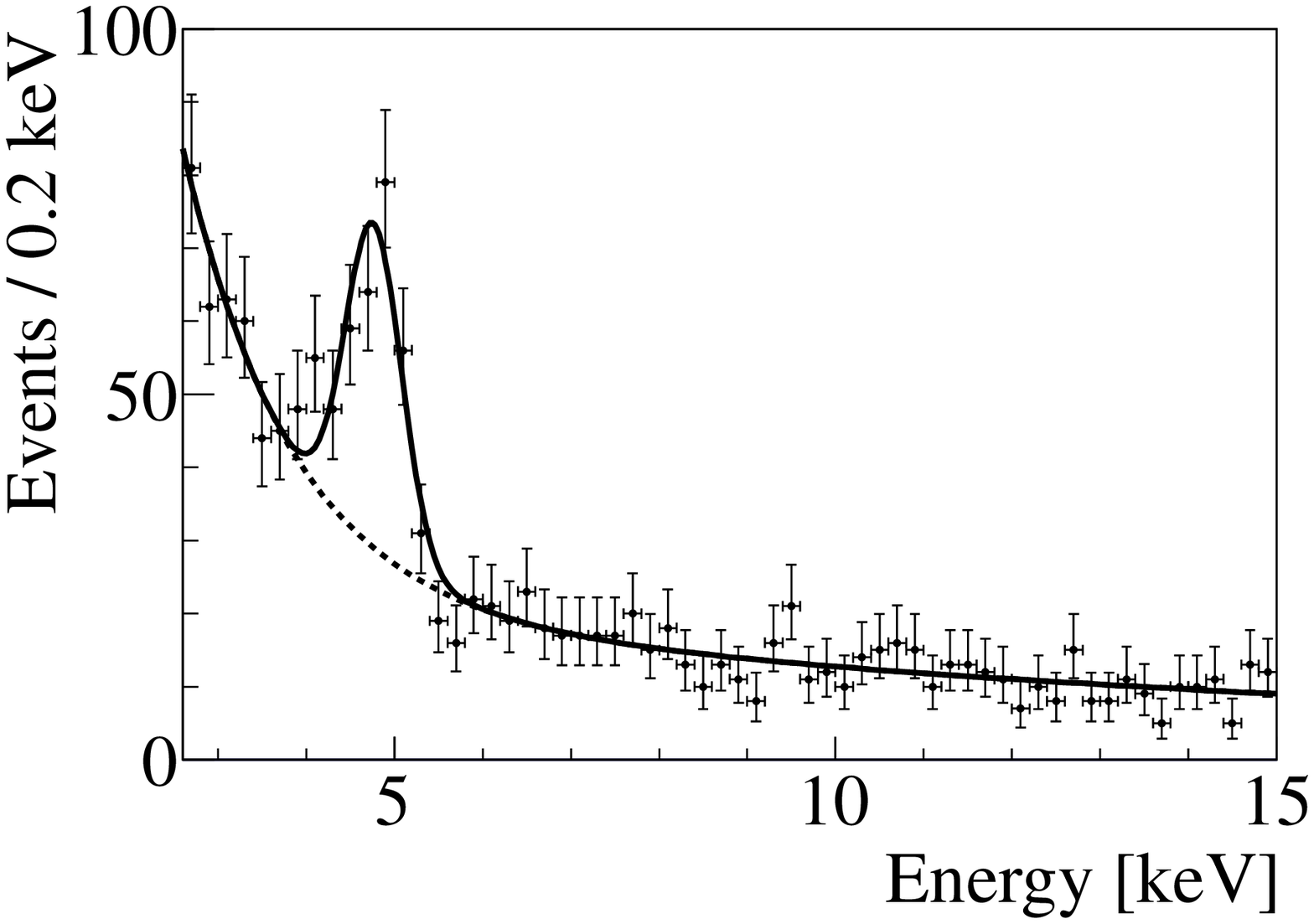}
\put(79,58){B4}
\end{overpic}
\includegraphics[width=0.218\textwidth,clip=true]{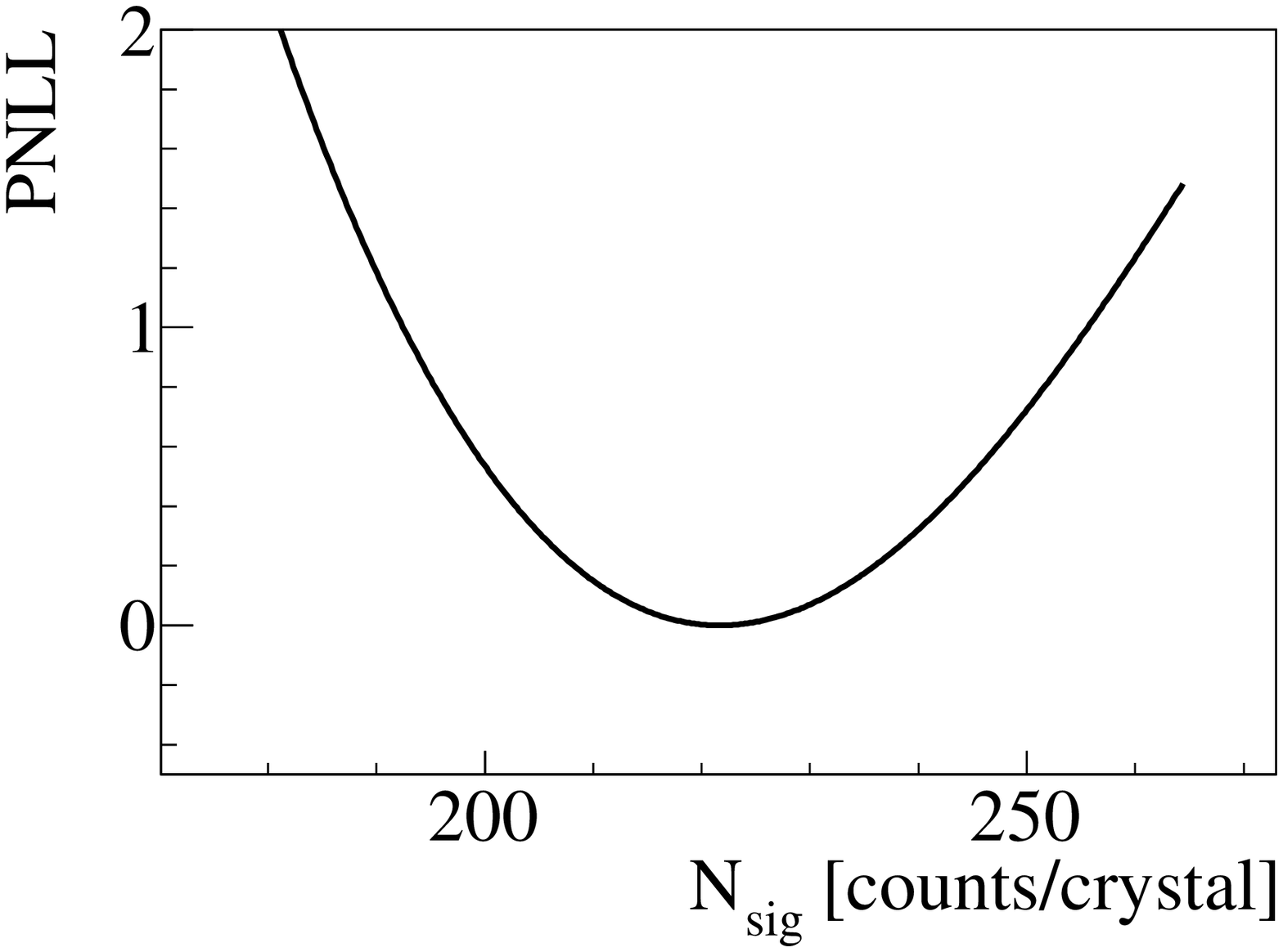} 
\caption{Best fits
at the $4.7\un{keV}$ line on the three bolometers with low threshold. 
On the bottom right, the combined profile negative log-likelihood
projected over the number of signal events.}
\label{fig:bestfit4keV}
\end{figure}

To estimate the intensity and the energy of the line from the
data, we performed a separate extended unbinned maximum-likelihood fit for each bolometer,
using a likelihood function constituted by a Gaussian plus two exponential
functions to reproduce the background.
We set the pulse shape cuts at the estimated optimal values.
Since the events in the 4.7\un{keV} peak are more plentiful
than in the Sb K peak at 30.5\un{keV}, the selection efficiencies
were recomputed, and confirmed to be equal to 1.
The obtained number of
events $N_{e}$ was then corrected taking into account the detection and cut efficiencies
and the expected background $N_{121}$, according to the equation
$N_{sig}=(N_{e}-N_{121})/(\epsilon_{D}\epsilon_{S})$. Best fits are
shown in Fig.~\ref{fig:bestfit4keV}. In Tab.~\ref{tab:fit4keV} we report
the summary of the peak parameters of each bolometer, also including the
estimated peaking background $N_{121}$. 

\begin{table}[tbp]
\begin{center}
\caption{Best fit results for the $4.7\un{keV}$ line and estimated peaking background
from $^{121}$Te. The error on the energy includes the systematic due to the calibration.}
\label{tab:fit4keV}
\begin{tabular}{|c|c|c|c|c|}
\hline
Bolometer& Energy &  FWHM&  $N_{sig}$ & $N_{121}$\\
 & [keV] & [keV] & [counts] & [counts] \\
\hline
B2 & $4.75\pm0.28$ & $0.57\pm0.11$  & $191^{+41}_{-31}$ & $6.1 \pm 1.3$\\
B3 & $4.66\pm0.28$ & $0.87\pm0.10$  & $255^{+51}_{-35}$ & $13.5 \pm 1.6$\\
B4 & $4.76\pm0.38$ & $0.75\pm0.12$  & $205^{+45}_{-34}$ & $7.5 \pm 1.4$\\
\hline
\end{tabular}
\end{center}
\end{table}

We combined the profile negative log-likelihoods of $N_{e}$ and $N_{121}$
of the three bolometers (Fig.~\ref{fig:bestfit4keV}) to compute the average rate. The estimated
value of the number of signal events is $223\pm 22\un{counts/crystal}$,
from which we evaluated the line intensity in \TEO\
to be $I=15.3\pm1.5\un{counts/day/kg}$.  The energy averaged over the
three bolometers is $4.72\pm0.18\un{keV}$.

\begin{figure}[htbp]
\centering \includegraphics[width=0.49\textwidth]{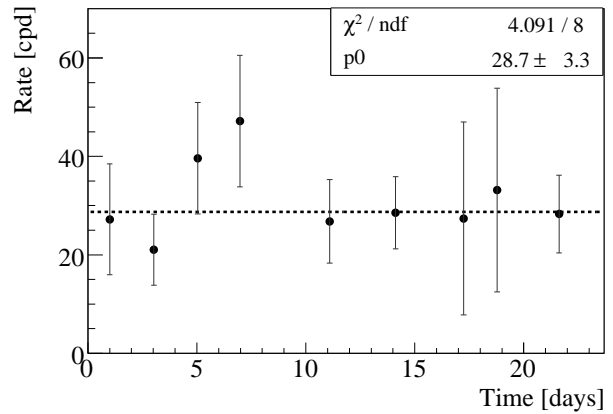}
\caption{Rate of the $4.7\un{keV}$ peak versus time in all
three CCVR2 bolometers summed together. No variation is appreciable.} \label{fig:LTe_rate_vs_time}
\end{figure}

Figure~\ref{fig:LTe_rate_vs_time} shows the peak rate along the duration
of the data-taking, which is in a very good agreement with a constant distribution, indicating
that this signal is not due to a short-living radioactive contamination. It also indicates
that any variation of the detection efficiency with time is negligible compared
to the statistical error.

To check the stability of the line over a longer period of time, we checked
its presence in the \CUORICINO\ data.  During the last two months of
operation of \CUORICINO, the data-acquisition system being developed for \CUORE\
(the same used for CCVR2), was run in parallel with the old system,
saving to disk the continuous stream of data.
The live-time of continuous data amounts to 47.5\un{days}.  We ran the
new trigger on these data but we were able to reach a threshold below
4\un{keV} only on 4 bolometers, because of the high vibrational noise
transmitted by the holder to the crystals (CCVR, \CUOREZ\ and \CUORE\
holders have been explicitly redesigned to lower the transmitted noise).
These 4 bolometers had smaller size
(3x3x6\un{cm^3}, 330\un{g}) than other \CUORICINO\ and \CUORE\ bolometers,
and featured higher signal to noise ratio.

\begin{figure}[t]
\centering \includegraphics[width=0.49\textwidth,clip=true]{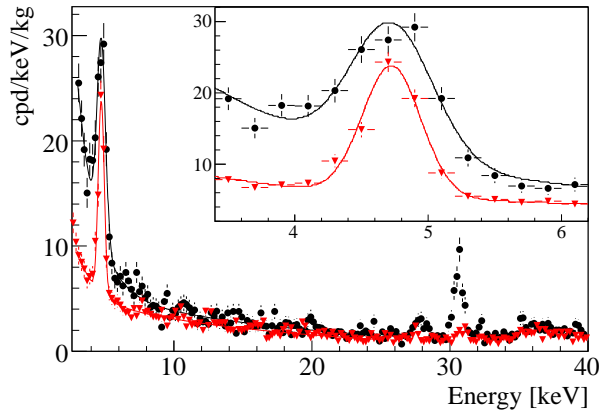}
\caption{Comparison of CCVR2 (black circles) and \CUORICINO\ (red triangles) data, with a zoom on the 4.7\un{keV}. Unlike
in CCVR2, \CUORICINO\ data are not corrected for the efficiencies and the calibration at low energies is not optimized.}
\label{fig:CCVRvscuoricinorate}
\end{figure}

The sum energy spectrum of the four bolometers is overlaid
to the sum energy spectrum of the CCVR2 bolometers in
Fig.~\ref{fig:CCVRvscuoricinorate}.  In \CUORICINO\ the
heater energy scan was not performed, therefore the spectrum shown is
not corrected for the detection efficiencies.
 As it can be seen from the
figure, the peak is very well visible, with a fitted energy
resolution of $0.52\pm0.04\un{keV~FWHM}$. The intensity and the
average energy are found to be similar to the CCVR2 ones, being
$(10.0\pm 0.6)/{\epsilon}\un{counts/day/kg}$ and $4.73\pm0.01\un{keV}$,
respectively. We also note that the 30.5\un{keV} peak disappeared,
because the $^{121m-121}$Te isotopes decayed away during the 5 years
of underground data taking. 

As the calibration was based on the $^{232}$Th source only, and given the
lack of low energy peaks, the accuracy of the calibration function at
low energies cannot be checked directly on \CUORICINO\ data. To have
an estimate of the calibration accuracy, during a new CCVR run we built a new setup in 
which a $^{55}$Fe source was deposited on the copper holder. The
detector was operated in the same setup used for the CCVR2 one, reaching
an energy resolution of 1.4\un{keV~FWHM} and an energy threshold of
around 3\un{keV}. To emulate the \CUORICINO\ conditions, the calibration
function was estimated using the peaks from the $^{232}$Th source only.
The X-rays produced by the $^{55}$Fe, with nominal energies between 5.888
and 6.490\un{keV}, resulted in detected energies that were shifted, on
average, by only $+(48\pm16)\un{eV}$ from their nominal values. Even
if the operating conditions of this setup were not identical to the
\CUORICINO\ ones, this shift can be taken as an indication of the
systematics associated to the calibration function.

\section{Sensitivity to WIMPs} 

WIMPs can couple to nucleons via both spin-independent and spin-dependent
(axial vector) interactions. Spin-independent scattering dominates when $A\ge30$ because, 
for low momentum transfer, it benefits from coherence across the nucleus~\cite{Jungman1996}. 
In addition, spin-dependent scattering is significant only on nuclei with an odd number of nucleons.
Since $99.8\%$ of Oxygen is constituted by $^{16}$O and $92\%$ of Tellurium is 
composed by even isotopes (mainly $^{130}$Te, $^{128}$Te and $^{126}$Te), \TEO\ will be most sensitive for spin-independent interacting WIMPs. It has to be remarked
that \TEO\ bolometers, unlike other bolometric detectors for Dark
Matter, do not discriminate the nuclear recoils induced by WIMPs from
the radioactive $\beta/\gamma$ background.  Nevertheless, the high mass
and low background achievable with these detectors make it possible to
search for an annual modulation of the counting rate.

Compared to \CUORICINO, the CCVR2 rate has the same behavior at energies
greater than 10\un{keV}, but is considerably higher at lower energies
(Fig.~\ref{fig:CCVRvscuoricinorate}).  
We are unable to explain this difference at present, however 
we expect that the \CUOREZ\ and \CUORE\ low energy background will not be 
higher than the CCVR2 one. All the materials used in detector construction, in fact, 
will be the same as those employed for CCVR2. 
Moreover, in the case of \CUOREZ, the bolometers will be operated in the same 
cryostat of \CUORICINO, which has lower radioactive contaminations compared to the one 
used to operate CCVR2.
To be conservative, we estimate the \CUOREZ\ and \CUORE\ sensitivity to WIMPs 
assuming the background rate of these experiments to be equal to the one 
measured on CCVR2. We assume that the noise will be under control and that all
bolometers will achieve a 3\un{keV} threshold. We focus on the energy
region between threshold and 25\un{keV}, featuring an observed background
counting rate ranging from about 25\un{cpd/keV/kg} to 2\un{cpd/keV/kg}.

\begin{figure}[t!]
\centering \begin{overpic}[width=0.49\textwidth]{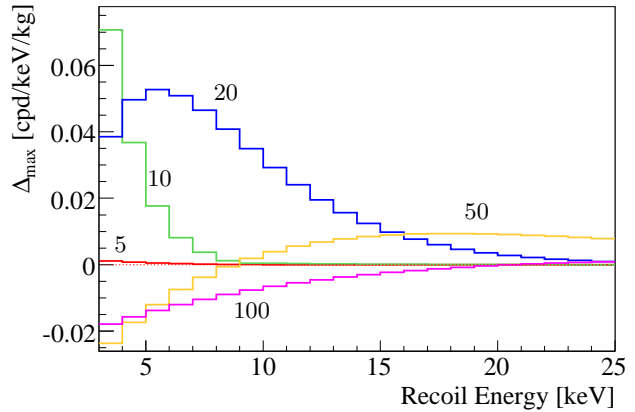}
\put(17,28){5}
\put(22,38){10}
\put(32,51){20}
\put(70,33){50}
\put(35,18){100}
\end{overpic}
\caption{WIMP simulated signal: difference between the 3-month integrated spectra across December the 2nd and
June the 2nd, for a WIMP cross section of $10^{-41}\un{cm^2}$ and masses between 5 and 100\un{GeV}.}
\label{fig:expmod}
\end{figure}

We perform toy \MC\ simulations generating background events from
the CCVR2 fit function shown in Fig.~\ref{fig:CCVRvscuoricinorate}, and WIMP events from the
predicted distribution described in Ref.~\cite{LewinSmith1996}, using
the following WIMP parameters: density $\rho_W=0.3\un{GeV/cm^3}$, average
velocity $v_0 = 220\un{km/s}$ and escape velocity from the Galaxy $v_{esc}
= 600\un{km/s}$.  The quenching factor (QF) of the interactions in \TEO\
is set to 1~\cite{Alessandrello:1997ca}.  We include the dependence
of the WIMP interaction rate on the time in the year, and estimate the
background+signal asymmetry subtracting the 3-month integrated spectrum
across  December the 2nd from the 3-month integrated spectrum across
June the 2nd.  The resulting differential spectrum is fitted with the
expected shape induced by the modulation, $H_1$ (examples are given
in Fig.~\ref{fig:expmod}), and with a flat line at zero counts, $H_0$.
The cross section in the toy simulation is lowered as long as, in a set
of experiments, the fit probability of the $H_1$ hypothesis
is greater than the $H_0$ one at least 90\% of the times. 

\begin{figure}[h!]
\centering \includegraphics[width=0.48\textwidth, clip=true]{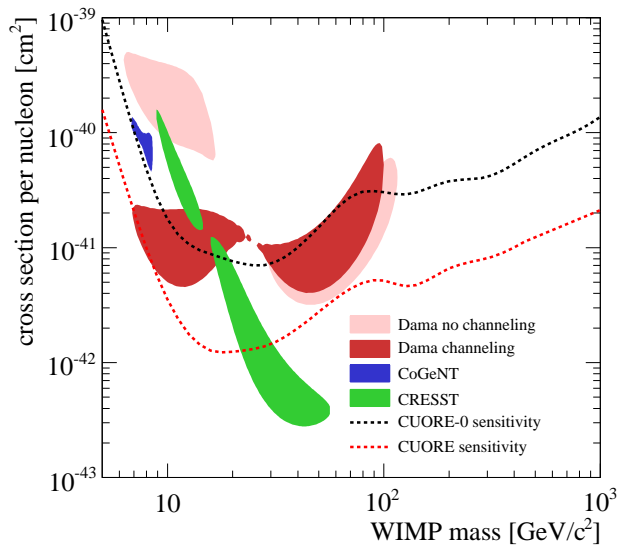}
\caption{90\% sensitivity to WIMP spin-independent scattering of \CUOREZ\ and \CUORE\ 
assuming a 3\un{keV} threshold for all detectors and the same background level of the CCVR2 detectors. Evidences of DAMA-$3\sigma$~\cite{Bernabei:2008yi},  CoGeNT-90\%~\cite{Aalseth:2011wp} and CRESST-$2\sigma$~\cite{Angloher:2011uu} are reported for comparison.}
\label{fig:exclusion}
\end{figure}

This procedure defines the cross section that could be sensed for a fixed
WIMP mass. The 90\% CL upper limit sensitivity to the cross-section as a function of the WIMP mass is reported in
Fig.~\ref{fig:exclusion} for 3-years of \CUOREZ\ data-taking and 5-years
of \CUORE. The comparison with other experiments shows that \CUOREZ\ could
test the indication of a $\sim 10\un{GeV}$ WIMP from the DAMA (no-channeling), 
CoGeNT and CRESST experiments, while \CUORE\ could completely test the DAMA
results, under the hypothesis that Dark Matter is purely made of
spin-independent interacting WIMPs.
We reiterate that because the quenching factor  for nuclear recoils compared to electron recoils in \TEO\ bolometers is 1, the 2-6 keV energy region of DAMA corresponds to 7-20 keV
assuming scattering on Na (QF=0.3) or 22-67 keV assuming scattering on I
(QF=0.9).  Therefore, in \TEO\ bolometers it will be possible to look at
lower energies and to study with larger detail the shape of the modulation
spectrum, thus providing new information to this complicated search.

\begin{acknowledgments}
The CUORE Collaboration thanks the Directors and Staff of the Laboratori Nazionali del Gran Sasso and the technical staffs of our Laboratories. 
This work was supported by the Istituto Nazionale di Fisica Nucleare (INFN); the Director, Office of Science, of the U.S. Department of Energy under Contract Nos. DE-AC02-05CH11231 and DE-AC52-07NA27344; the DOE Office of Nuclear Physics under Contract Nos. DE-FG02-08ER41551 and DEFG03-00ER41138; the National Science Foundation under Grant Nos. NSF-PHY-0605119, NSF-PHY-0500337, NSF-PHY-0855314, and NSF-PHY-0902171; the Alfred P. Sloan Foundation; and the University of Wisconsin Foundation.
\end{acknowledgments}

\cleardoublepage

\bibliography{main}

\begin{thebibliography}{35}%
\makeatletter
\providecommand \@ifxundefined [1]{%
 \@ifx{#1\undefined}
}%
\providecommand \@ifnum [1]{%
 \ifnum #1\expandafter \@firstoftwo
 \else \expandafter \@secondoftwo
 \fi
}%
\providecommand \@ifx [1]{%
 \ifx #1\expandafter \@firstoftwo
 \else \expandafter \@secondoftwo
 \fi
}%
\providecommand \natexlab [1]{#1}%
\providecommand \enquote  [1]{``#1''}%
\providecommand \bibnamefont  [1]{#1}%
\providecommand \bibfnamefont [1]{#1}%
\providecommand \citenamefont [1]{#1}%
\providecommand \href@noop [0]{\@secondoftwo}%
\providecommand \href [0]{\begingroup \@sanitize@url \@href}%
\providecommand \@href[1]{\@@startlink{#1}\@@href}%
\providecommand \@@href[1]{\endgroup#1\@@endlink}%
\providecommand \@sanitize@url [0]{\catcode `\\12\catcode `\$12\catcode
  `\&12\catcode `\#12\catcode `\^12\catcode `\_12\catcode `\%12\relax}%
\providecommand \@@startlink[1]{}%
\providecommand \@@endlink[0]{}%
\providecommand \url  [0]{\begingroup\@sanitize@url \@url }%
\providecommand \@url [1]{\endgroup\@href {#1}{\urlprefix }}%
\providecommand \urlprefix  [0]{URL }%
\providecommand \Eprint [0]{\href }%
\@ifxundefined \urlstyle {%
  \providecommand \doi  [0]{\begingroup \@sanitize@url \@doi}%
  \providecommand \@doi [1]{\endgroup \@@startlink {\doibase
  #1}doi:\discretionary {}{}{}#1\@@endlink }%
}{%
  \providecommand \doi  [0]{doi:\discretionary{}{}{}\begingroup
  \urlstyle{rm}\Url }%
}%
\providecommand \doibase [0]{http://dx.doi.org/}%
\providecommand \Doi [0]{\begingroup \@sanitize@url \@Doi }%
\providecommand \@Doi  [1]{\endgroup\@@startlink{\doibase#1}\@@Doi}%
\providecommand \@@Doi [1]{#1\@@endlink}%
\providecommand \selectlanguage [0]{\@gobble}%
\providecommand \bibinfo  [0]{\@secondoftwo}%
\providecommand \bibfield  [0]{\@secondoftwo}%
\providecommand \translation [1]{[#1]}%
\providecommand \BibitemOpen [0]{}%
\providecommand \bibitemStop [0]{}%
\providecommand \bibitemNoStop [0]{.\EOS\space}%
\providecommand \EOS [0]{\spacefactor3000\relax}%
\providecommand \BibitemShut  [1]{\csname bibitem#1\endcsname}%
\bibitem {Arnaboldi:2003tu}
 C.~Arnaboldi\ \emph {et~al.},
 Astropart.\ Phys.\ {\bf 20}, 91 (2003).
%
\bibitem {ACryo}
 C.~Arnaboldi\ \emph {et~al.},
 Nucl.\ Instrum.\ Meth.\ A\ {\bf 518}, 775 (2004).
%
\bibitem {Redshaw:2009zz}
M.~Redshaw,\ B.~J.~Mount,\ E.~G.~Myers\ and\ F.~T.~Avignone,
Phys.\ Rev.\ Lett.\ {\bf 102}, 212502 (2009).
%
\bibitem {scielzo09}
N.~D.~Scielzo\ \emph {et~al.},
Phys.\ Rev.\ C\ {\bf 80}, 025501 (2009).
%
\bibitem {Rahaman:2011zz}
S.~Rahaman\ \emph {et~al.},
Phys.\ Lett.\ B\ {\bf 703}, 412 (2011).
%
\bibitem {Andreotti:2010vj}
E.~Andreotti\ \emph {et~al.},
Astropart.\ Phys.\ {\bf 34}, 822 (2011).
%
\bibitem {Andreotti:2011in}
E.~Andreotti\ \emph {et~al.},
Phys.\ Rev.\ C\ {\bf 85}, 045503 (2012).
%
\bibitem{Bertone:2004pz}
  G.~Bertone,\ D.~Hooper\ and\ J.~Silk,
      Phys.\ Rept.\  {\bf 405},  279 (2005).
%
\bibitem{Steigman:1984ac} 
  G.~Steigman\ and\ M.~S.~Turner,
      Nucl.\ Phys.\ B\ {\bf 253}, 375 (1985).
%
\bibitem{Goodman:1984dc} 
  M.~W.~Goodman\ and\ E.~Witten,
      Phys.\ Rev.\ D\ {\bf 31}, 3059 (1985).
%
\bibitem{Drukier:1986tm} 
  A.~K.~Drukier,\ K.~Freese\ and\ D.~N.~Spergel,
      Phys.\ Rev.\ D\ {\bf 33}, 3495 (1986).
%
\bibitem{pdg2012}
J.~Beringer\ \emph{et~al.} (Particle\ Data\ Group), 
Phys.\ Rev.\ D\ {\bf 86}, 010001 (2012).
%
\bibitem {DiDomizio:2010ph}
S.~Di~Domizio,\ F.~Orio\ and\ M.~Vignati,
JINST\ {\bf 6}, P02007 (2011).
%
\bibitem {Gatti:1986cw}
E.~Gatti\ and\ P.~F.~Manfredi,
Riv.\ Nuovo\ Cimento\ {\bf 9}, 1 (1986).
%
\bibitem {Radeka:1966}
V.~Radeka\ and\ N.~Karlovac,
Nucl.\ Instrum.\ Methods\ {\bf 52}, 86 (1967).
%
\bibitem {Alessandria:2011vj}
F.~Alessandria\ \emph {et~al.},
Astropart.\ Phys.\ {\bf 35}, 839 (2012).
%
\bibitem {wang}
N.~Wang\ \emph {et~al.},
Phys.\ Rev.\ B\ {\bf 41}, 3761 (1990).
%
\bibitem {Itoh}
K.~M.~Itoh\ \emph {et~al.},
Appl.\ Phys.\ Lett.\ {\bf 64}, 2121 (1994).
%
\bibitem {Mott:1969}
N.~F.~Mott,
Philos.\ Mag.\ {\bf 19}, 835 (1969).
%
\bibitem {Itoh:1996}
K.~M.~Itoh\ \emph {et~al.},
Phys.\ Rev.\ Lett.\ {\bf 77}, 4058 (1996).
%
\bibitem {AProgFE}
C.~Arnaboldi\ \emph {et~al.},
IEEE\ T.\ Nucl.\ Sci.\ {\bf 49}, 2440 (2002).
%
\bibitem {stabilization}
A.~Alessandrello\ \emph {et~al.},
Nucl.\ Instrum.\ Meth.\ A\ {\bf 412}, 454 (1998).
%
\bibitem {Arnaboldi:2003yp}
C.~Arnaboldi,\ G.~Pessina\ and\ E.~Previtali,
IEEE\ T.\ Nucl.\ Sci.\ {\bf 50}, 979 (2003).
%
\bibitem {Andreotti:2012zz}
  E.~Andreotti\ \emph {et~al.},
Nucl.\ Instrum.\ Meth.\ A\ {\bf 664}, 161 (2012).
%
\bibitem {Pirro:2006mu}
S.~Pirro,
Nucl.\ Instrum.\ Meth.\ A\ {\bf 559}, 672 (2006).
%
\bibitem {Arnaboldi:2006mx}
C.~Arnaboldi,\ G.~Pessina\ and\ S.~Pirro,
Nucl.\ Instrum.\ Meth.\ A\ {\bf 559}, 826 (2006).
%
\bibitem {Arnaboldi:2004jj}
C.~Arnaboldi\ \emph {et~al.},
Nucl.\ Instrum.\ Meth.\ A\ {\bf520}, 578 (2004).
%
\bibitem {Carrettoni:2011rn}
M.~Carrettoni\ and\ M.~Vignati,
JINST\ {\bf 6}, P08007 (2011).
%
\bibitem {Ohya20101619}
S.~Ohya,
Nucl.\ Data\ Sheets\ {\bf 111}, 1619 (2010).
%
\bibitem {ohya2004nuclear}
S.~Ohya,
Nucl.\ Data\ Sheets\ {\bf 102}, 547 (2004).
%
\bibitem {watt1962search}
D.~Watt\ and\ R.~Glover,
Philos.\ Mag.\ {\bf 7}, 105 (1962).
%
\bibitem {zuber:2003}
D.~Munstermann\ and\ K.~Zuber,
J.\ Phys.\ G\ Nucl.\ Partic.\ {\bf 29}, B1 (2003).
%
\bibitem {PhysRevC.67.014323}
A.~Alessandrello\ \emph {et~al.},
Phys.\ Rev.\ C\ {\bf 67}, 014323 (2003).
%
\bibitem {allison2006geant4}
J.~Allison\ \emph {et~al.},
IEEE\ T.\ Nucl.\ Sci.\ {\bf 53}, 270 (2006).
%
\bibitem {Jungman1996}
G.~Jungman,\ M.~Kamionkowski\ and\ K.~Griest,
Phys.\ Rept.\ {\bf 267}, 195 (1996).
%
\bibitem {LewinSmith1996}
J.~D.~Lewin\ and\ P.~F.~Smith,
Astropart.\ Phys.\ {\bf 6}, 87 (1996).
%
\bibitem {Alessandrello:1997ca}
A.~Alessandrello\ \emph {et~al.},
Phys.\ Lett.\ B\ {\bf 408}, 465 (1997).
%
\bibitem {Bernabei:2008yi}
R.~Bernabei\ \emph {et~al.},
Eur.\ Phys.\ J.\ C\ {\bf 56}, 333 (2008).
%
\bibitem {Aalseth:2011wp}
C.~Aalseth\ \emph {et~al.},
Phys.\ Rev.\ Lett.\ {\bf 107}, 141301 (2011).
%
\bibitem {Angloher:2011uu}
G.~Angloher\ \emph {et~al.},
Eur.\ Phys.\ J.\ C\ {\bf 72}, 1971 (2012).
%
\end{thebibliography}%

\end {document}